\documentclass[superscriptaddress,twocolumn,10pt, prb, preprintnumbers,amsmath,amssymb,colorlinks,floatfix, longbibliography]{revtex4-2}

\usepackage{graphicx}
\usepackage{dcolumn}
\usepackage{bm}
\usepackage{array}
\usepackage{amsmath,bm}
\newcolumntype{P}[1]{>{\centering\arraybackslash}p{#1}}

\usepackage{lipsum}

\usepackage{nameref}
\usepackage{varioref}
\usepackage{hyperref}
\usepackage{cleveref}
\usepackage{comment}

\newcommand{\tbiropur}{Tb$_{2}$Ir$_{2}$O$_{7}$}
\newcommand{\tbiro}{Tb$_{2+x}$Ir$_{2-x}$O$_{7-y}$}

\newcommand{\euiropur}{Eu$_{2}$Ir$_{2}$O$_{7}$}
\newcommand{\airo}{$A_{2}$Ir$_{2}$O$_7$}

\begin{document}

\title{Spin dynamics and possible topological magnons in non-stoichiometric pyrochlore iridate \tbiropur\ studied by RIXS}

\author{Q.~Faure}
\email[Corresponding author.~Electronic address: ]{quentin.faure@cea.fr}
\affiliation{ESRF, The European Synchrotron, 71 Avenue des Martyrs, CS40220, 38043 Grenoble Cedex 9, France}
\affiliation{Laboratoire L\'eon Brillouin, CEA, CNRS, Universit\'e Paris-Saclay, CEA-Saclay, 91191 Gif-sur-Yvette, France}
\affiliation{London Centre for Nanotechnology and Department of Physics and Astronomy,University College London, Gower Street, London WC1E6BT, UK}

\author{A.~Toschi}
\affiliation{ESRF, The European Synchrotron, 71 Avenue des Martyrs, CS40220, 38043 Grenoble Cedex 9, France}
\affiliation{Institut de Physique, \'Ecole Polytechnique F\'ed\'erale de Lausanne (EPFL), CH-1015 Lausanne, Switzerland}

\author{J.~R.~Soh}
\affiliation{A*STAR Quantum Innovation Centre (Q.InC), Institute for Materials Research and Engineering (IMRE), Agency for Science, Technology and Research (A*STAR), 2 Fusionopolis Way, 08-03 Innovis 138634, Republic of Singapore}
\affiliation{Institut de Physique, \'Ecole Polytechnique F\'ed\'erale de Lausanne (EPFL), CH-1015 Lausanne, Switzerland}

\author{E.~Lhotel}
\affiliation{Institut N\'eel, CNRS and Universit\'e Grenoble Alpes, 38000 Grenoble, France}

\author{B.~Detlefs}
\affiliation{ESRF, The European Synchrotron, 71 Avenue des Martyrs, CS40220, 38043 Grenoble Cedex 9, France}

\author{D.~Prabhakaran}
\affiliation{Department of Physics, University of Oxford, Clarendon Laboratory, Oxford, OX1 3PU, United Kingdom}

\author{D.~F.~McMorrow}
\affiliation{London Centre for Nanotechnology and Department of Physics and Astronomy,University College London, Gower Street, London WC1E6BT, UK}

\author{C.~J.~Sahle}
\affiliation{ESRF, The European Synchrotron, 71 Avenue des Martyrs, CS40220, 38043 Grenoble Cedex 9, France}

\date{\today}

\begin{abstract}
We report a resonant Inelastic X-ray scattering study on a single crystal of a non-stoichiometric pyrochlore iridate \tbiro\ ($x \simeq 0.25$) that magnetically orders at $T_{\rm{N}}\simeq 50$~K. 
We observe a propagating gapped magnon mode at low energy, and model it using a Hamiltonian consisting of a Heisenberg exchange [$J = 16.2(9)$~meV] and Dzyaloshinskii-Moriya interactions [$D = 5.2(3)$~meV], which shows the robustness of interactions despite Tb-\textit{stuffing} at the Ir-site. Strikingly, the ratio $D/J = 0.32(3)$ supports possible non-trivial topological magnon band crossing. This material may thus host coexisting fermionic and bosonic topology, with potential for manipulating electronic and magnonic topological bands thanks to the $d-f$ interaction.    


\end{abstract}

\maketitle

Iridium based compounds hosting a $j_{\rm{eff}} =1/2$ ground state have attracted a lot of interest due to the delicate interplay between large spin-orbit coupling (SOC), moderate electronic correlations and crystal field that can give rise to novel exotic electronic and magnetic states~\cite{Krempa2014, Schaffer2016, Rau2016}. In particular, topological phases of matter are predicted to arise such as topological insulators, axion insulators and Weyl semi-metal (WSM) phases~\cite{Binghai2017}.

\begin{figure}[t]
\centering{\includegraphics[width=\linewidth]{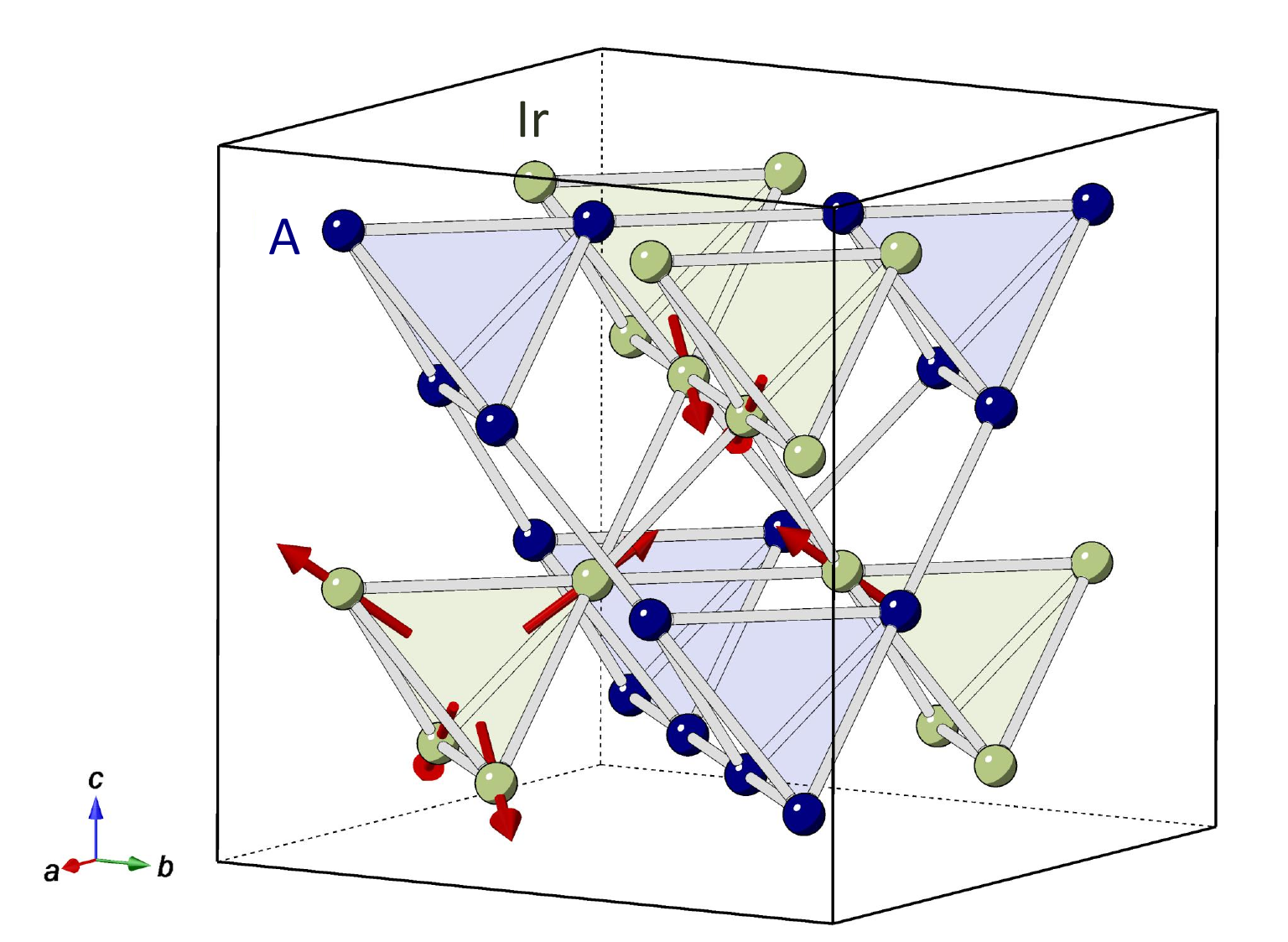}}
\caption{Pyrochlore structure of \airo\ with two interpenetrating rare earth (blue spheres) and iridium sublattices (yellow spheres) forming corner shared tetrahedra. The AIAO magnetic order corresponds to spins (red arrows) pointing either within or out of a given tetrahedra. Such a magnetic configuration is depicted only on two tetrahedra for better clarity.}
\label{Figure1}
\end{figure}

In this context, pyrochlore iridates with formula \airo\ ($A$ = Y, or rare earth Lu-Pr) have been extensively studied as these systems were the first potential candidates to realize WSM phases driven by magnetic ordering~\cite{Pesin2010, Yang2010, Wan2011, Witczak2012, Go2012, Shinaoka2015}. These compounds crystallize in the Fd$\bar{3}$m cubic structure (space group No.~227) consisting of two interpenetrating rare earth and iridium pyrochlore sublattices (Fig.~\ref{Figure1}) that are connected via a $d-f$ interaction making them a fertile playground to discover novel phases and to study frustrated magnetism~\cite{Pesin2010, Chen2012, Krempa2014, Schaffer2016}. The whole family of pyrochlore iridates exhibits metal-to-insulator transition (MIT) when cooling down in temperature, concomitant with magnetic ordering (except for $A$ = Pr)~\cite{Matsuhira2011}. The iridium sublattice was shown to order in an all-in-all-out (AIAO) magnetic structure (Fig.~\ref{Figure1}) compatible with the realization of a WSM state~\cite{Witczak2012}, using various techniques such as $\mu$-SR, neutron powder diffraction and Resonant Elastic X-rays scattering~\cite{Zhao2011, Tomiyasu2012, Sagayama2013, Disseler2014, Lefrancois2015, Clancy2016, Guo2016}. While the existence of the WSM phase in pyrochlore iridates is still theoretically debated~\cite{Wan2011, Witczak2012, Go2012, Chen2012, Shinaoka2015, Jan2017, Wang2017, Zhang2017}, experimental signatures of fermionic band-topology in these compounds were observed, e.g. through optical THz spectroscopy~\cite{Sushkov2015} or magneto-resistivity measurements where novel field induced WSM phases were observed in Nd$_2$Ir$_2$O$_7$~\cite{Ueda2015, Tian2016, Ueda2017}. Studies of thin films also showed topological properties of the WSM phase by breaking cubic symmetry through epitaxial strain~\cite{Yang2014, Liu2021, Ghosh2022}. 

Interestingly, topological properties in pyrochlore iridates extend beyond fermions. Indeed, similarly to its ferromagnetic counterpart found in the pyrochlore compound Lu$_2$V$_2$O$_7$ where a magnon Hall effect~\cite{Onose2010, Mena2014} was observed, recent studies proposed that pyrochlore iridates could host topological bosons, i.e. topologically non-trivial magnon bands depending on the ratio of the Dzyaloshinskii–Moriya interaction (DMI) $D$ and Heisenberg interaction $J$~\cite{Laurell2017, Hwang2020}. More precisely, if $D/J > 0.28$, magnon band crossing is predicted to occur close to the $\Gamma$ point. This makes pyrochlore iridates unique systems where both fermionic and bosonic band topology coexist and hence potential platforms to realize topological electronic and magnonic devices~\cite{McClarty2022}. A better understanding of the spin dynamics in these compounds is thus crucial, yet challenging. The Ir$^{4+}$ cation is indeed a strong neutron-absorber and posseses a small magnetic moment ($m_{\rm{Ir}} \simeq 0.3$~$\mu_{\rm{B}}$), making inelastic neutron scattering unsuccessful up to now. Resonant inelastic x-ray scattering (RIXS), on the other hand, is a powerful tool, since it allows to probe spin and orbital dynamics in iridium compounds due to resonant processes enhancing magnetic scattering cross sections~\cite{Ament2011}. It is worth noting that while the spin dynamics in pyrochlore iridates has been explored in systems with non-magnetic rare earth ions~\cite{Donnerer2016, Chun2018, Nguyen2021}, little attention has been given to those with magnetic rare earth ions, where the $d-f$ interaction may modify electronic and magnonic band structure and thus drive novel topological phases~\cite{Chen2012, Krempa2014, Jacobsen2020}. In particular, \tbiropur\ is of interest owing to the relatively large moments of the Tb$^{3+}$ cations ($m_{\rm{Tb}} \simeq 5.2$~$\mu_{\rm{B}}$)~\cite{Guo2016} that could easily couple to a magnetic field and thus potentially drive exotic phase transitions, as observed in Nd$_2$Ir$_2$O$_7$~\cite{Ueda2015, Ueda2017, Tian2016}.


In this letter, we report a study using RIXS at the Ir $L_3$-edge of a non-stoichiometric single crystal of \tbiro\ ($x \simeq 0.25$) in which the transition temperature is strongly suppressed down to 50~K (instead of 130~K for stoichiometric powder \tbiropur). Analysis of the orbital $dd$ intra-t$_{2g}$ excitations yields spin-orbit coupling $\xi = 0.40(1)$~eV and trigonal distortion of the IrO$_6$ octahedra $\Delta=0.52(2)$~eV. We observe a propagating gapped magnetic excitation and analyze the dispersion spectra with a linear spin-wave model. This allows us to extract Heisenberg exchange interaction $J = 16.2(9)$~meV and DMI $D = 5.2(3)$~meV, leading to the ratio $D/J = 0.32(3)$. Our findings demonstrate that interactions remain robust despite Tb-\textit{stuffing} on Ir-site. More strikingly, the found $D/J = 0.32(3)$ ratio supports the possible presence of topological magnons in our \textit{stuffed} sample of \tbiropur. This material may thus host coexisting fermionic and bosonic topology, with potential for manipulating electronic and magnonic topological bands via the $d-f$ interaction.

\begin{figure}[!h]
\centering{\includegraphics[width=\linewidth]{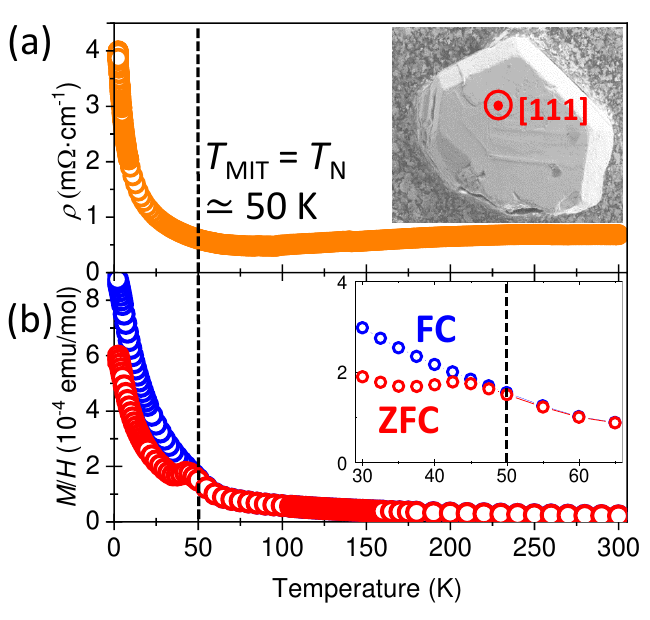}}
\caption{(a) Resistivity $\rho$ vs temperature. Inset shows a photograph of the sample of \tbiro\ of size $L \times l \times h \simeq 0.3 \times 0.3 \times 0.2$~mm$^3$. The sample has a hexagonal shape with the [111] direction perpendicular to its surface. (b) ZFC-FC magnetization data. The inset shows a zoom around the transition temperature $T_{N} = T_{\rm{MIT}} \simeq 50$~K where a bifurcation between the ZFC-FC curves is observed.}
\label{Figure2}
\end{figure}

\tbiro\ single crystals were flux-grown as described in \cite{Millican2007} (see inset of Fig.~\ref{Figure2}). Electron dispersive X-ray spectroscopy (EDX) was used to probe the cation ratio, which revealed an excess of Tb ions of $x \simeq 0.25$ with no reliable estimation of the oxygen content (details in the Supplemental Material~\cite{supmat} that includes Refs.~\cite{Matsuhira2011, Donnerer2016, Moretti2013, Dynaflow2016, Liu2012, Sala2014, Chun2018, Elhajal2005, Yadav2018, Hwang2020, Nguyen2021, Toth2015}). The crystal structure probed by single crystal X-ray diffraction gives a Fd${\bar 3}$m pyrochlore structure with lattice parameter $a = 10.26$~\AA, slightly larger than previously reported in \tbiropur\ powder samples~\cite{Lefrancois2015, Guo2016}.

Electrical and magnetic macroscopic properties were characterized by four-probe resistivity and magnetization measurements using PPMS and MPMS Quantum Design apparatus respectively. Resistivity vs temperature measurements show a MIT concomitant with a magnetic ordering of the iridium sublattice at $T_{\rm{N}} = T_{\rm{MIT}} \simeq 50$~K, as observed by the bifurcation between the zero-field-cooled (ZFC) and field-cooled (FC) magnetization data (Fig.~\ref{Figure2}). This magnetic ordering is further confirmed by resonant x-ray magnetic diffraction measurements (details in Supplemental Material~\cite{supmat}). This value is much lower than for pure \tbiropur\ ~\cite{Matsuhira2011}, indicating that \textit{stuffing} tends to suppress the onset of magnetic order. This is associated with a resistivity which below $T_{\rm{MIT}}$ is  3-4 orders of magnitude lower than for \tbiropur~\cite{Matsuhira2011}. These observations are consistent with previous studies on {\it stuffed} pyrochlore iridates, and especially \tbiro\ ($x = 0.18$)~\cite{Donnerer2019}, where off-stoichiometry was shown to favor metallic or semi-metallic behaviors ~\cite{Ishikawa2012, Clancy2016, Telang2018}.

\begin{figure}[!h]
\centering{\includegraphics[width=\linewidth]{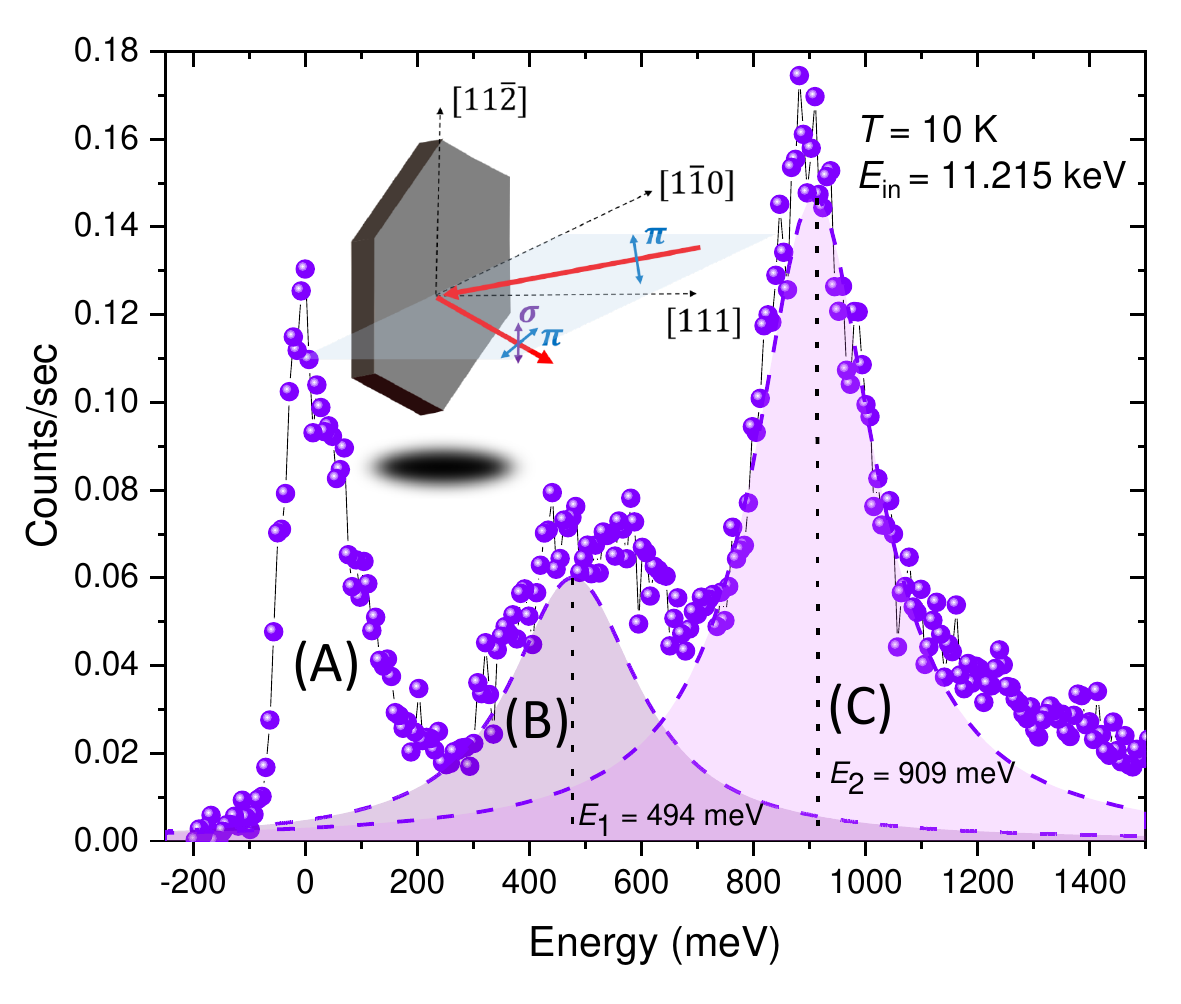}}
\caption{High resolution RIXS spectrum collected at $T = 10$~K and $\mathbf{Q} = (7.5,7.5,7.5)$. The incident energy is set to $E{_{in}}=11.215$~keV. Full purple circles denote the experimental data. An incoherent elastic contribution with low-energy excitations (A) is observed. Two higher energies $dd$ intra $t_{2g}$ excitations (B) and (C) are visible and are fitted with Lorentzian functions (dashed purple and pink curves) leading to $E_1 = 494(7)$~meV and $E_2 = 909(3)$~meV. The inset shows the horizontal geometry used in the RIXS experiment with the scattering plane defined by the $[111]$ and $[1\bar{1}0]$ directions of the sample.}
\label{Figure3}
\end{figure}

A RIXS experiment at the Iridium $L_3$-edge was performed on the beamline ID20 of the European Synchrotron Radiation Facility with an overall energy resolution of 25~meV~\cite{Moretti2013} (details in the Supplemental Material~\cite{supmat}). The scattering plane and incident photon polarization were both horizontal in the laboratory frame, i.e. $\pi$ incident polarization was used (inset in Fig.~\ref{Figure3}). In order to enhance the intra-t$_2g$ magnetic and orbital signal, we tuned $E_{\rm{in}}$ to a resonant condition, specifically the Ir $L_3$-edge in this case. We thus set $E_{\rm{in}} = 11.215$~keV in the present study.

Fig.~\ref{Figure3} depicts a high-energy resolution RIXS spectrum measured at $T = 10$~K. Within the spectrum, three features are observed. First, there is an incoherent elastic contribution followed by a low-energy inelastic signal (A) down to 300 meV. Then, two higher energy excitations (B) and (C) are observed around 0.5 and 1 eV. By fitting the latter two excitations with Lorentzian functions, their energies are found to be $E_1 = 494(7)$~meV and $E_2 = 909(3)$~meV respectively. Note that that these excitations do not disperse, as previously observed in other pyrochlore iridates \cite{supmat}. \textit{Ab-initio} calculations~\cite{Hozoi2014} along with RIXS measurements~\cite{Clancy2016} have shown that these two excitations correspond to $dd$ excitations within the t$_{2g}$ manifold. The presence of these two features is explained by the combination of spin-orbit coupling $\xi$ and trigonal distortion $\Delta$ of the IrO$_6$ octahedra, which can be described by 
a single-ion model for Ir$^{4+}$ in a 5d t$_{2g}$ basis subjected to trigonal distortion~\cite{Liu2012, Hozoi2014}:  
\begin{equation*}
H = \xi \bm{L} \bm{S} - \Delta L_z^2 
\end{equation*}
where $\xi$ denotes the spin-orbit coupling and $\Delta$ the trigonal distortion with the quantization axis $z$ corresponding to the three-fold rotation axis of the octahedron. From an effective model, one can extract $\xi$ and $\Delta$ from the $dd$ excitations with energies $E_1$ and $E_2$ with the following relations~\cite{Liu2012, Hozoi2014}: $\lambda = 2(2 E_1 - E_2)/(3-\delta)$, $\Delta = \lambda \delta/2$ with $\delta = -b-\sqrt{b^2-9}$, $b = (1+3 a^2)/(1-a^2)$ and $a = E_2/(E_2-2 E_1)$.
We thus extracted the following values: $\xi = 0.40(1)$~eV and $\Delta = 0.52(2)$~eV, which are consistent with the values found for other pyrochlore iridates~\cite{Clancy2016, Hozoi2014}. We now turn to the analysis of the low-energy excitations below 300 meV.

\begin{figure*}[t!]
\centering{\includegraphics[width=\linewidth]{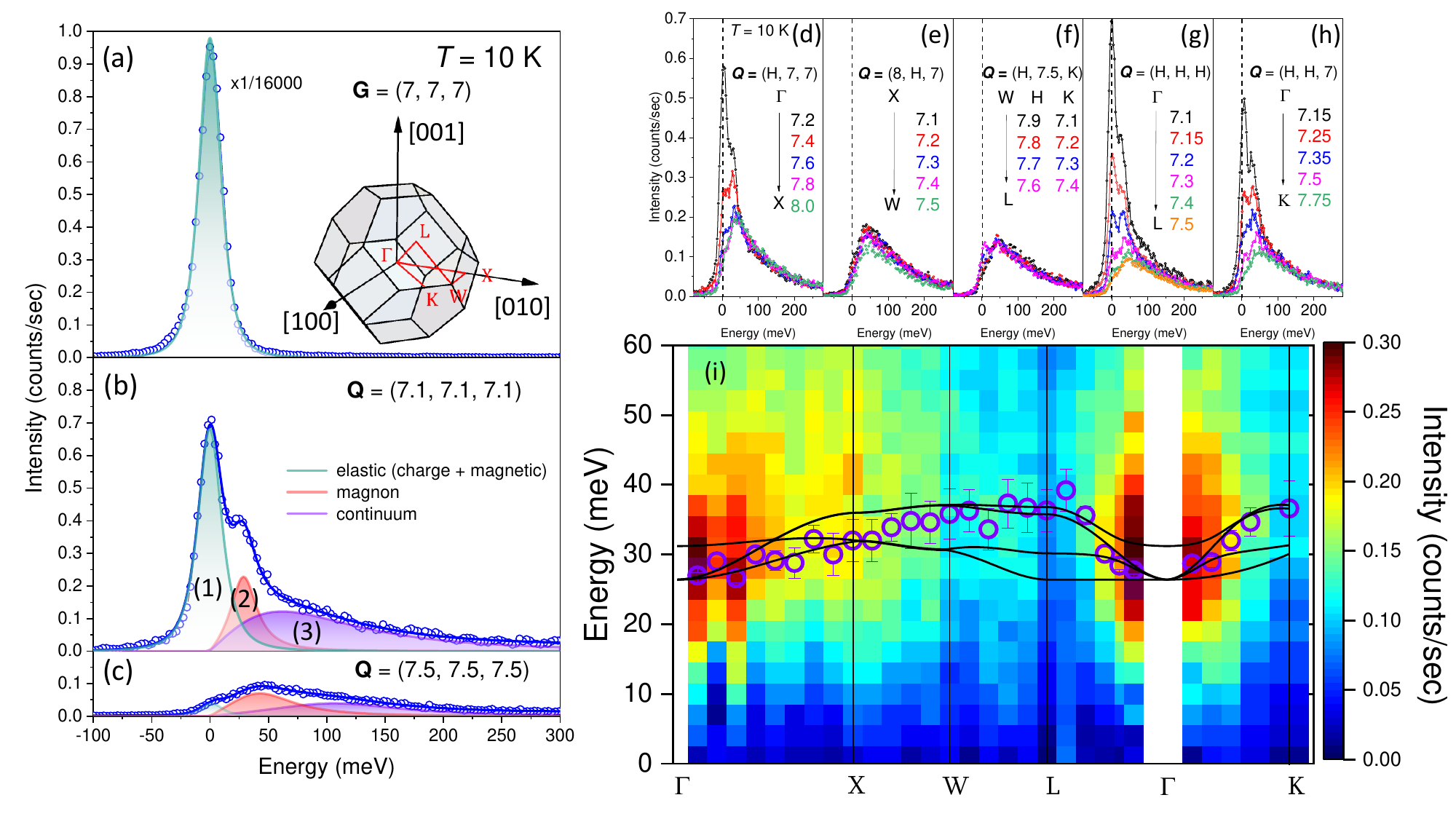}}
\caption{(a-c) High-resolution RIXS spectra collected at $T = 10$~K and $\bm{Q} = G = (7, 7, 7)$ (a), $\bm{Q} = (7.1, 7.1, 7.1)$ and $\bm{Q} = (7.5, 7.5, 7.5)$. The intensity in (a) is reduced by a factor of 16000 for comparison to other spectra. The elastic line (1) (green curve) is fitted using a Pseudo-Voigt function. The spectra in (b-c) show a sharp inelastic feature (2) that corresponds to the magnon (red solid line) and an inelastic continuum (3) (purple solid line). Open blue circles correspond to raw data. The global fitting curve (blue solid line) corresponds to the sum of those three features. Inset: first Brillouin zone (BZ) of the face cubic centered lattice with high-symmetry points and directions. (d-h) RIXS spectra recorded at $T = 10$~K and along the high-symmetry directions within the BZ: $\mathrm{\Gamma-X}$ (d), $\mathrm{X-W}$ (e), $\mathrm{W-L}$ (f), $\mathrm{\Gamma-L}$ (g) and $\mathrm{\Gamma-K}$ (h). (i) Dispersion spectra obtained from (d-h). The elastic line has been substracted for more clarity. The open purple circles denote the energy of the magnetic excitations extracted from the fits of the spectra in (d-h), and the associated error bars. 
The four black lines denote the calculated four non-degenerate magnon dispersions using SpinW~\cite{Toth2015} with $J = 16.2$~meV and $D = 5.2$~meV, leading to the best agreement with the experimental dispersion.}
\label{Figure4} 
\end{figure*}

Fig.~\ref{Figure4}(a, b, c) show the RIXS spectra measured at $T = 10$~K at the zone-center $\bm{Q} = \bm{\mathrm{G}} = (7, 7, 7)$, $\bm{Q} = (7.1, 7.1, 7.1)$ and  $\bm{Q} = (7.5, 7.5, 7.5)$. Three main features appear: (1) an incoherent elastic contribution centered at zero energy-loss; (2) a sharp dispersing mode around 20-50 meV and (3) a broad continuum of excitations from 50 to 300 meV. The dispersing sharp mode (2) corresponds to a magnetic excitation as it is not present at room temperature \cite{supmat}. The continuum of excitations (3) does not seem to disperse and its nature is still unclear at present. Indeed, this broad continuum, previously seen in other RIXS studies of pyrochlore iridates~\cite{Clancy2016, Chun2018, Donnerer2016}, has a broad linewitdh and is temperature independent. Its origin has been attributed to be a particle-hole continuum, e.g. inter-band excitations, or incoherent multimagnetic excitations~\cite{Chun2018}. 


To fit our data and in order to follow precisely the dispersion of the magnetic excitation, we used a damped harmonic-oscillator model, also used in Ref.~\cite{Chun2018} (note that fits using a Lorentzian curve for the single-magnon gave the same results onto its energy position): 
\begin{equation*}
S(\bm{q},\omega) = \frac{A_{\bm{q}}}{(1-e^{-\omega/T})}\left[\frac{\gamma_{\bm{q}}}{(\omega-\omega_{\bm{q}})+\gamma_{\bm{q}}^2}-\frac{\gamma_{\bm{q}}}{(\omega+\omega_{\bm{q}})+\gamma_{\bm{q}}^2}\right]
\label{DOH}
\end{equation*}
where $\omega_{\bm{q}}$ is the energy peak position, $\gamma_{\bm{q}}$ the peak width and $A_{\bm{q}}$ the overall amplitude. The broad inelastic continuum (3) was fitted using the same function while the elastic background (1) was fitted using a Pseudo-Voigt function. Fig.~\ref{Figure4}(d-h) show scans along high-symmetry directions within the Brillouin-zone (see inset of Fig.~\ref{Figure4}(a)): $\mathrm{\Gamma-X}$, $\mathrm{X-W}$, $\mathrm{W-L}$, $\mathrm{\Gamma-L}$ and $\mathrm{\Gamma-K}$ respectively. By fitting the whole set of data with the above fitting function, we were able to precisely obtain the full dispersion of the magnetic mode as depicted in Fig.~\ref{Figure4}(i). Our data indicate a spin gap of 27~meV and a dispersion width of 12 meV. This gapped spectrum is consistent with an AIAO magnetic order~\cite{Chun2018, Donnerer2016}. The magnon mode is strongly dispersing along $\mathrm{\Gamma-L}$ and $\mathrm{\Gamma-K}$ while almost not dispersing along $\mathrm{X-W}$ and $\mathrm{W-L}$. Interestingly, it weakly disperses along $\mathrm{\Gamma-X}$, in contrast to what was reported in Sm$_2$Ir$_2$O$_7$ and Eu$_2$Ir$_2$O$_7$~\cite{Donnerer2016, Chun2018}. The intensity of the magnon mode is strongly suppressed when moving away from the $\mathrm{\Gamma}$ point, consistent with what was observed previously in other pyrochlore iridates~\cite{Chun2018, Donnerer2016}.

To model our data, we use a linear spin-wave approach including the minimal spin Hamiltonian for AIAO pyrochlore iridates  (details in Supplemental Material~\cite{supmat}): 
\begin{equation*}
H = \sum_{<i,j>} J \bm{S}_{i} \bm{S}_{j} + \bm{D}_{ij}\cdot (\bm{S}_{i} \times  \bm{S}_{j})
\label{Hamiltonian}
\end{equation*} 
where $J$ corresponds to the Heisenberg exchange interaction between nearest neighbors $<i,j>$ and $\bm{D}_{ij}$ the antisymetric DMI. Using SpinW~\cite{Toth2015}, the experimental magnon dispersion was fitted through linear spin wave calculations (see Fig.~\ref{Figure4}(i)), yielding to $J = 16.2(9)$~meV and $D = |\bm{D}_{ij}| = 5.2(3)$~meV and thus to a ratio $D/J = 0.32(3)$. The calculated curves capture very well the dispersion along $\mathrm{\Gamma-K}$, $\mathrm{\Gamma-L}$ and $\mathrm{L-W}$. However discrepancies appear along $\mathrm{X-W}$ and only the lowest energy magnon band captures the dispersion of the magnon mode along $\mathrm{\Gamma-X}$. These discrepancies could be explained by the fact that a spin-wave approach assumes strong electron correlations $U$ while pyrochlore iridates were shown to be in a more itinerant regime~\cite{Chun2018}. To go further, a more sophisticated approach such as RPA of the dynamical spin susceptibility would be necessary~\cite{Lee2013}. The presence of Ir$^{4+}$ vacancies should also be taken into account but it is expected to mainly introduce a broadening of the magnon bands. Given the low energy resolution of RIXS, this thus not appears crucial in the present analysis, and one can consider the $J$ and $D$ values as average values over the random distribution of Ir$^{4+}$ vacancies in \tbiro. 

Interestingly, the $D/J = 0.32(3)$ value found here supports non-trivial topological magnon band crossing as pointed out by recent theoretical studies~\cite{Son2019, Hwang2020, Nguyen2021}. Indeed, for $D/J > 0.28$ magnon band crossing occurs along $\Gamma-\rm{X}$, which is reproduced by our spin-wave calculations. 
The $D/J$ ratio depends on the Ir-O-Ir bond angle, which itself depends linearly on the ionic radius of the $A^{3+}$ cation in \airo. Hence $D/J$ is expected to decrease smoothly within the Lu-Nd series of magnetically AIAO ordered pyrochlore iridates. More specifically, $J$ and $D$ decreases and increases respectively with the size of the ionic radius. Recent Raman studies showed that in the case of Y$^{3+}$ owing a small ionic radius, Y$_2$Ir$_2$O$_7$ hosts $D/J = 0.68$ with $D = 9.0$~meV and $J = 15.1$~meV. Consistently, RIXS studies on \euiropur\ and Sm$_2$Ir$_2$O$_7$, thus with larger ionic radius, found a ratio of $D/J = 0.26, 0.18$~\cite{Chun2018, Donnerer2016, Hwang2020}. Our findings are consistent with this dependence of $J$ and $D$ with the ionic radius, i.e. Tb$^{3+}$ has an intermediate size and is placed between Y$^{3+}$ and Eu$^{3+}$, Sm$^{3+}$~\cite{Nguyen2021}. These results nevertheless contrast from the ones calculated for \tbiropur\ through quantum chemistry calculations (QCC) which proposed a scenario where $D = 5$~meV~$\gg$~$J = -1.5$~meV~\cite{Yadav2018}. This may be due to a different value of the Ir-O-Ir angle used in the calculations compared to the material~\cite{Nguyen2021}.
Finally, we demonstrate that the exchange couplings are robust despite \textit{stuffing}. This could be explained by the large extended 5$d$ orbitals of the Ir$^{4+}$ cations and the three dimensional character of this system. The precise effect of \textit{stuffing}, and how the exchange couplings are affected in comparison with pure \tbiropur\, are nevertheless unclear and we let this open question for future studies. Note that recent results in Nd-based pyrochlores have shown that magnetic interactions are little affected even in the presence of strong disorder and high off-stoichiometry~\cite{Leger2024}.


In summary, we investigated a \textit{stuffed} sample of \tbiro\ ($x \simeq 0.25$) by means of RIXS. The low-energy spectra exhibit a dispersive magnon gapped excitation consistent with AIAO order. A linear spin wave modelling of the experimental magnon dispersion allowed us to extract the Heisenberg interaction $J = 16.2(9)$~meV and the DMI $D = 5.2(3)$~meV leading to $D/J = 0.32(3)$. These values are consistent with the trend of $J$ and $D$ that depends on Ir-O-Ir bond-angle and hence ionic radius. Strikingly the ratio $D/J$ found here supports topological magnon band crossing in \textit{stuffed} \tbiropur. This establishes this material as an interesting platform where both fermionic and magnonic topology coexist and where electronic and magnonic bands could be modulated thanks to the $d-f$ exchange. We also demonstrate that magnetic interactions are robust despite the presence of \textit{stuffing}. We hope our findings will motivate future complementary measurements, e.g. Raman scattering, thermal Hall measurements or RIXS with higher energy-resolution to demonstrate the presence of non-trivial topological magnon bands in \textit{stuffed} and pure \tbiropur.

\acknowledgments

The authors would like to thank M.~Moretti~Sala, S.~Petit and P.A.~McClarty for fruitful discussions. We thank F.~Gerbon for technical support during the resonant inelastic x-rays scattering (RIXS) measurements on ID20 at ESRF and E.~Pachoud for for technical support on EDX measurements performed at Institut Néel. We acknowledge ESRF for allocation of RIXS beamtime on ID20. D.~Phrabakaran acknowledges the Engineering and Physical Sciences Research Council (EPSRC), UK grant number  EP/T028637/1 and the Oxford-ShanhagaiTech collaboration project for financial support.
   
\bibliography{biblio.bib}

\end{document}


\title{Supplementary material : Spin dynamics and possible topological magnons in non-stoichiometric pyrochlore iridate \tbiropur\ studied by RIXS}

\author{Q.~Faure}
\email[Corresponding author.~Electronic address: ]{quentin.faure@cea.fr}
\affiliation{ESRF, The European Synchrotron, 71 Avenue des Martyrs, CS40220, 38043 Grenoble Cedex 9, France}
\affiliation{Laboratoire Léon Brillouin, CEA, CNRS, Université Paris-Saclay, CEA-Saclay, 91191 Gif-sur-Yvette, France}
\affiliation{London Centre for Nanotechnology and Department of Physics and Astronomy,University College London, Gower Street, London WC1E6BT, UK}

\author{A.~Toschi}
\affiliation{ESRF, The European Synchrotron, 71 Avenue des Martyrs, CS40220, 38043 Grenoble Cedex 9, France}

\author{J.R Soh}
\affiliation{A*STAR Quantum Innovation Centre (Q.InC), Institute for Materials Research and Engineering (IMRE), Agency for Science, Technology and Research (A*STAR), 2 Fusionopolis Way, 08-03 Innovis 138634, Republic of Singapore}
\affiliation{Institut de Physique, École Polytechnique Fédérale de Lausanne (EPFL), CH-1015 Lausanne, Switzerland}

\author{E. Lhotel}
\affiliation{Institut Néel, CNRS and Université Grenoble Alpes, BP166, F-38042 Grenoble Cedex 9, France}

\author{B.~Detlefs}
\affiliation{ESRF, The European Synchrotron, 71 Avenue des Martyrs, CS40220, 38043 Grenoble Cedex 9, France}

\author{D. Prabhakaran}
\affiliation{Department of Physics, University of Oxford, Clarendon Laboratory, Oxford, OX1 3PU, United Kingdom}

\author{D.~F.~McMorrow}
\affiliation{London Centre for Nanotechnology and Department of Physics and Astronomy,University College London, Gower Street, London WC1E6BT, UK}

\author{C.~J.~Sahle}
\affiliation{ESRF, The European Synchrotron, 71 Avenue des Martyrs, CS40220, 38043 Grenoble Cedex 9, France}

\date{\today}

\maketitle

\section{Energy dispersive X-rays spectroscopy}

\begin{figure}[!h]
\centering{\includegraphics[width=0.8\linewidth]{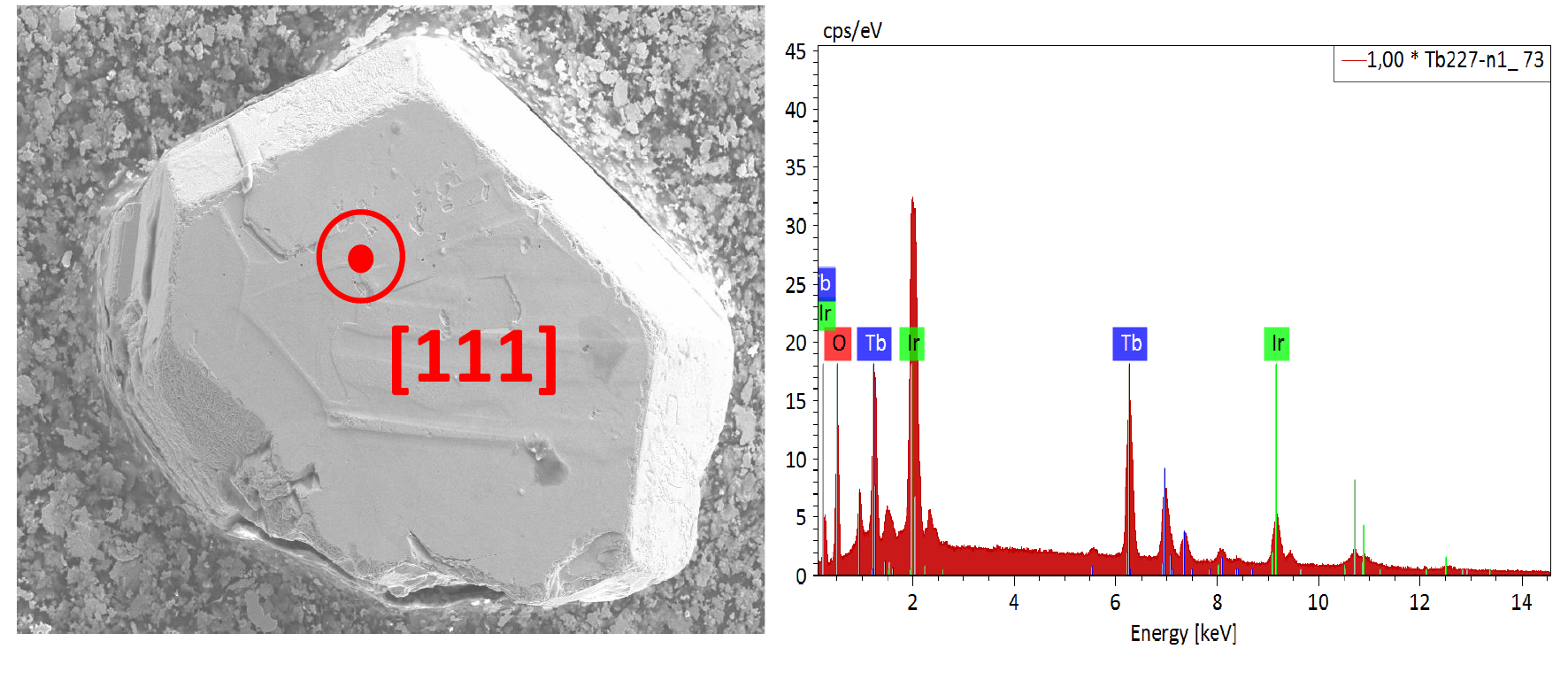}}
\caption{Characterization. Left: Photograph of the sample of \tbiro\ of size $L \times l \times h \simeq 0.3 \times 0.3 \times 0.2$~mm$^3$. Right: Electron dispersive X-ray analysis spectrum obtained from the sample.}  
\label{Figure0}
\end{figure}

Our first step involved the characterizion of our \tbiro sample. The sample exhibits a hexagonal shape with the $[111]$ direction perpendicular to its surface (see Fig.~\ref{Figure0}). 

To get an estimation of the off stoichiometry of our sample, we performed energy dispersive X-ray spectroscopy (EDX) using scanning electron microscopy (SEM) with a FESEM ZEISS Ultra$+$ equipped with a BRUKER-SSD. From EDX spectra (see Fig.~\ref{Figure0}), we found an off stoichiometry Tb~$:$~Ir $= 0.56:0.44$, without estimation of the oxygen content.

\begin{figure}[h!]
\centering{\includegraphics[width=0.4\linewidth]{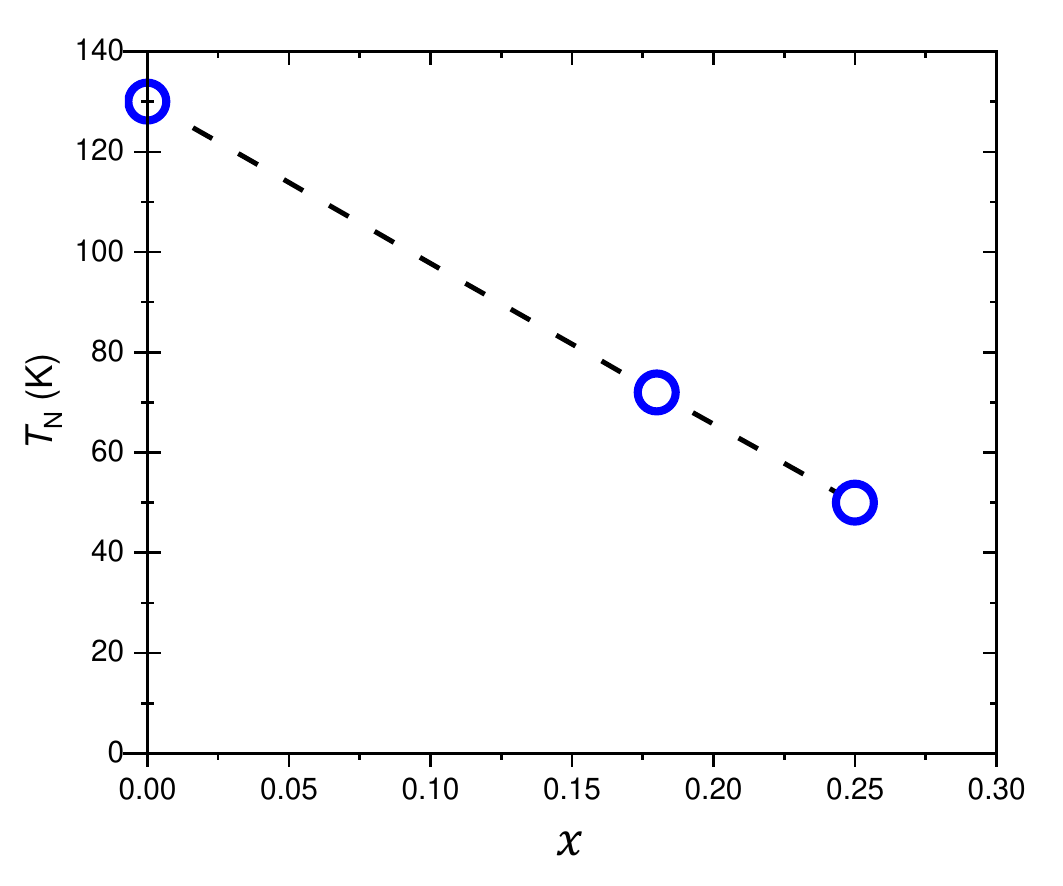}}
\caption{Evolution of the transition temperature $T_{\rm{N}}$ as a function of \textit{stuffing} $x$ for \tbiro. Dashed lines are guide to the eye.}
\label{Figure1}
\end{figure}

This ratio is consistent with the transition temperature of $T_{\rm{N}} = T_{\rm{MIT}}\simeq50$~K. Indeed, the transition temperature $T_{\rm{N}}$ follows a linear dependence as a function of $x$ for \tbiro\ as shown in Fig.~\ref{Figure1}. Table~\ref{table1} summarizes the different values of $T_{\rm{N}}$ for different $stuffed$ samples of \tbiro.

\begin{table}[h!]
  \begin{center}
    \caption{Different values magnetic ordering temperature $T_{\rm{N}}$ for different \textit{stuffed} samples of \tbiro\ with corresponding references.}
    \label{table1}
    \begin{tabular}{c|c|c}
      $x$ & $T_{\rm{N}}$ (K) & Ref. \\
      \hline
      0 & 130 & Ref.~\cite{Matsuhira2011}  \\
      0.18 & 71 & Ref.~\cite{Donnerer2019} \\
      0.25 & 50 & This work\\
    \end{tabular}
  \end{center}
\end{table}

\section{Magnetic resonant diffraction}

\begin{figure}[h!]
\centering{\includegraphics[width=0.7\linewidth]{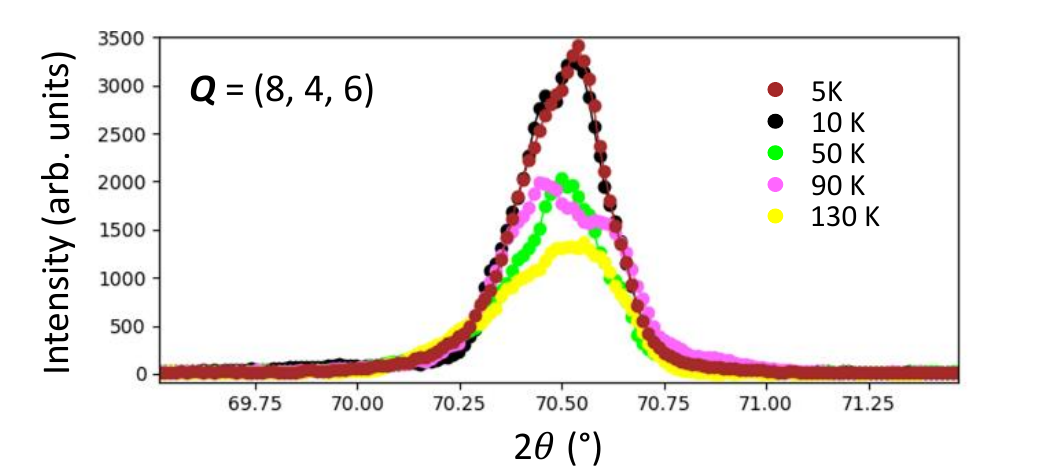}}
\caption{Temperature evolution of the magnetic Bragg peak $\bm{Q} = (8, 4, 6)$ measured by resonant elastic x-ray scattering.}
\label{Figure2}
\end{figure}

We measured $\theta-2 \theta$ scans for different temperature onto the $\bm{Q} = (8, 4, 6)$ reflection as function of temperature as shown in Fig.~\ref{Figure2}. The integrated intensity is maximum at 5 and 10~K and is reduced to become constant above $T = 50$~K, hence signaling the magnetic transition at this temperature. The remaining signal comes from anisotropic tensor scattering (ATS). However, we did not measured the azimutal dependence of the magnetic peak. Nevertheless, the gapped magnetic spectra observed in our experiment is consistent with the all-in-all-out magnetic order (see Ref.~\cite{Donnerer2016}) and can be interpreted with this ground state. Indeed, we assumed the all-in-all-out order as the starting point of our spin wave calculations; model that allows to reproduce the experimental magnon dispersion as explained later.

\section{Incident energy vs RIXS energy loss map}

The high-energy resolution RIXS experiment at the Iridium $L_3$-edge was performed on the beamline ID20 of the European Synchrotron Radiation Facility with an overall energy resolution of 25~meV~\cite{Moretti2013}. The incident beam was monochromatized using a Si(884) backscattering channel-cut crystal. The spectrometer in Rowland geometry ($R = 2$~m) was equipped with Si(553) diced analyzer crystals with an asymmetry offset of 5.05$^\circ$ allowing to reach the Si(884) reflection. The scattering plane and incident photon polarization were both horizontal in the laboratory frame, i.e. $\pi$ incident polarization was used. The sample was cooled down to 10~K using a closed flux helium cryostat~\cite{Dynaflow2016}. The RIXS process at the Ir $L_3$-edge is a second-order process consisting of two dipole transitions ($2p \rightarrow 5d$ followed by $5d \rightarrow 2p$). In order to enhance the intra-t$_{2g}$ magnetic and orbital signal, it is necessary to tune the incident energy $E_{\rm{in}}$ to a resonant condition, specifically the Ir $L_3$-edge in this case. Measuring a low-energy resolution incident energy vs energy loss RIXS map is thus a natural first step as it allows to find the resonant condition as explained below.  

\begin{figure}[!h]
\centering{\includegraphics[width=0.5\linewidth]{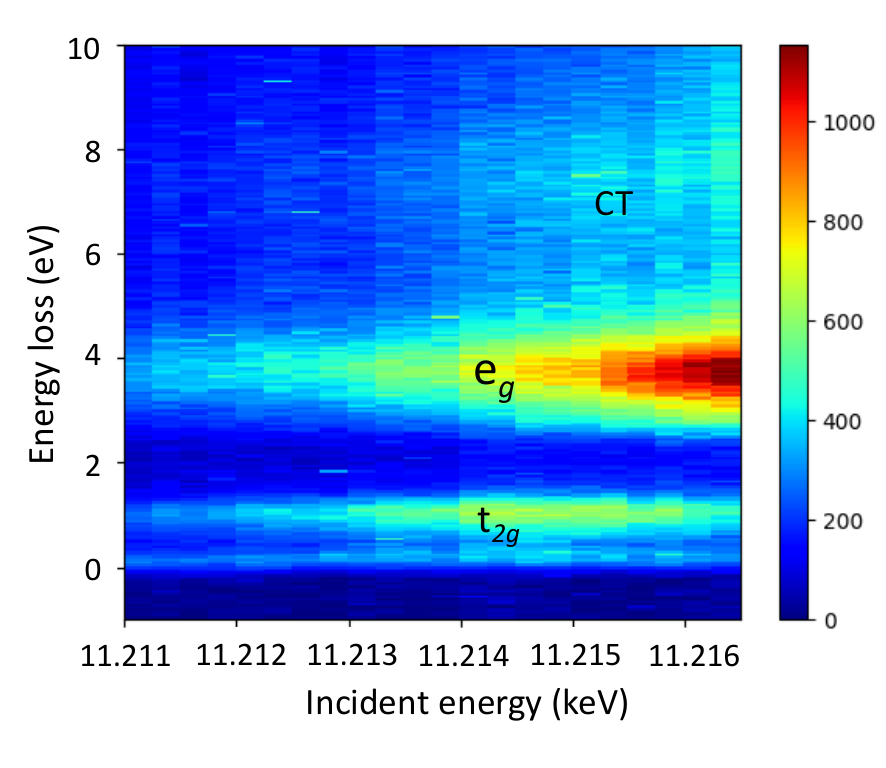}}
\caption{Low-energy resolution RIXS map measured at T = $20$~K and $\bm{Q} = (7.5, 7.5, 7.5)$ showing incident energy dependence of RIXS energy loss features in \tbiro. The observed energy loss features correspond to excitations within the t$_{2g}$ manifold (so called $dd$ intra t$_{2g}$ excitations), crystal field excitation to the e$_g$ manifold (t$_{2g}$ to e$_g$) and charge transfer (CT) excitations.}  
\label{Figure3}
\end{figure}

Before measuring high-resolution RIXS spectra, our initial step involved the measurement of a resonance map to determine the most suitable incident energy $E_{\rm{in}}$. To achieve this, a Si(311) channel cut post-monochromator was used, resulting in an overall energy resolution of 320 meV. The incident energy $E_{in}$ was varied within the range of 11.2111 to 11.2165 keV. Fig~\ref{Figure3} shows the RIXS resulting map measured at $T = 20$~K. 

This RIXS map shows the same features to those observed in previous studies on iridium oxides~\cite{Liu2012, Sala2014, Donnerer2016}: one feature below 2~eV energy transfer corresponding to intra-t$_{2g}$ excitations  and a broad feature around 4~eV energy loss corresponding to t$_{2g}$ to e$_g$ excitations. Another feature above 6~eV appears and is attributed to charge-transfer excitations. The two-step RIXS scattering process for these excitations, from initial to intermadiate to final state, can be written as follow: 

\begin{equation*}
5d(\rm{t}^5_{2g}) \rightarrow 
\left\lbrace
	\begin{aligned}
	&2\rm{p}^5_{3/2}5d(\rm{t}^6_{2g}) \rightarrow 
	\left\lbrace
	5d(\rm{t}^{5\star}_{2g}) \quad \text{dd intra-t$_{2g}$} \right.
    \\
	&2\rm{p}^5_{3/2}5d(\rm{e}^1_g) \rightarrow
		\left\lbrace 
		 \begin{aligned}
		 		&5d(\rm{t}^{5\star}_{2g}) \quad \text{t$_{2g}$ to e$_g$}\\
				&5d(t^4_{2g})5d(\rm{e}^1_{g})5d(\rm{e}^1_g)\underline{\textit{L}} \quad \text{CT}
		 \end{aligned}
		 \right.
	\end{aligned}
	\right.
\label{RIXSprocess}
\end{equation*} 
where the $\star$ denotes an excited state configuration and $\underline{L}$ a ligand hole. 

For high-energy resolution RIXS spectra and to enhance the magnetic signal, we chose to set the incident energy to $E_{\rm{in}} = 11.215$~keV which corresponds to the energy resonance of the intra-t$_{2g}$ excitations.

\section{Temperature dependence of the magnon}

\begin{figure}[!h]
\centering{\includegraphics[width=\linewidth]{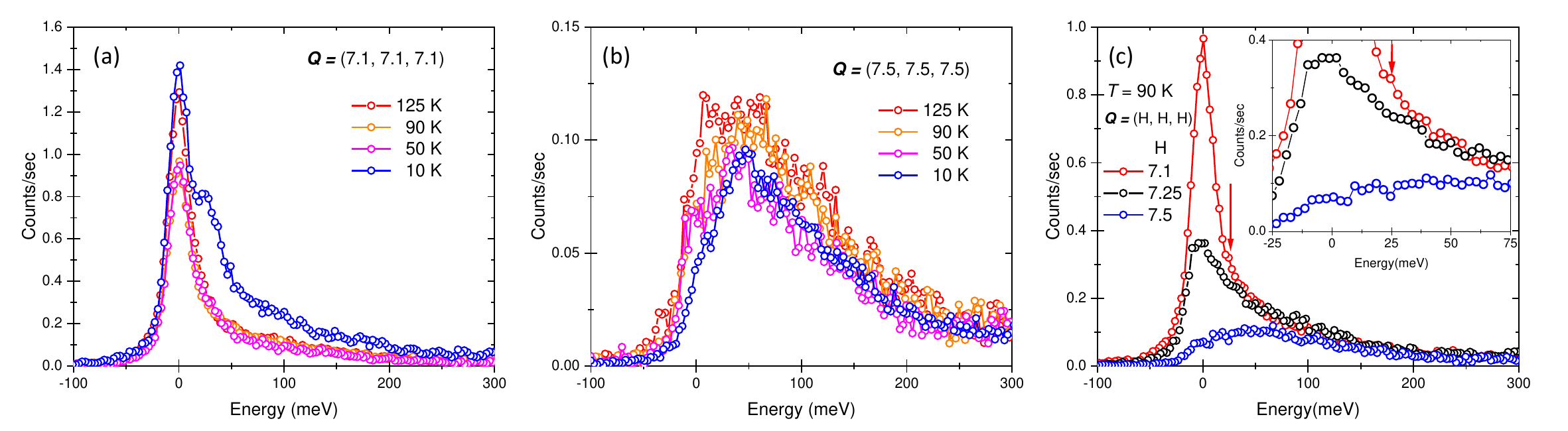}}
\caption{Temperature dependence of the magnon mode. (a-b) Energy scans for various temperature at $\bm{Q} = (7.1, 7.1, 7.1)$ and $\bm{Q} = (7.5, 7.5, 7.5)$. Open circles denote experimental data. Lines are guides to the eye. (c) Dispersion along $\mathrm{\Gamma-X}$ at $T = 90$ K. Open circles show experimental data. The red arrow points out the presence of a spin gap. The inset shows a zoom of the present energy scans.}  
\label{Figure4}
\end{figure}

Figs.~\ref{Figure4}(a-b) show the temperature dependence of the magnon mode for $\bm{Q} = (7.1, 7.1, 7.1)$ and $\bm{Q} = (7.5, 7.5, 7.5)$ respectively. The intensity of the sharp magnon mode present at $T = 10$~K is strongly suppressed for temperatures above $T_{\rm{MIT}} = T_{\rm{N}} = 50$~K. The continuum of excitations (feature (3) in the main text) is temperature independent as shown in Fig.~\ref{Figure4}(b). Fig.~\ref{Figure4}(c) shows the dispersion along $\mathrm{\Gamma-X}$ measured at $T = 90$~K. The magnon mode survives above $T_{\rm{N}} = 50$~K with strong reduction of its intensity, consistent with what was reported in Ref.~\cite{Chun2018}, i.e. that pyrochlore iridates cannot be described with a local moment model. Surprisingly though, the spin gap in our case seems to survive in contrast with Ref.~\cite{Chun2018} where it vanishes in the case of \euiropur. The detailed study of the temperature dependence of the spin gap is beyond the scope of this paper and further studies are required to solve this issue. 

\section{Spin-wave calculations}

The general spin-Hamiltonian proposed for pyrochlore iridates can be written as follows: 

\begin{equation*}
H = \sum_{<i,j>} J \bm{S}_i \bm{S}_j + \bm{D}_{ij}\cdot(\bm{S}_i\times\bm{S}_j) + \bm{S}_i \bm{\mathcal{A}}_{ij} \bm{S}_j
\label{Hamiltonianbis}
\end{equation*}
where $J$ is the Heisenberg exchange parameter, $\bm{D}_{ij}$ the antisymmetric Dzyaloshinskii-Moriya interaction and $\bm{\mathcal{A}}_{ij}$ the anisotropic exchange second-rank traceless tensor interaction. Based on symmetry analysis~\cite{Elhajal2005, Yadav2018, Hwang2020}, $\bm{D}_{ij}$ and $\bm{\mathcal{A}}_{ij}$ are written in the local $(x,y,z)$ basis where $x$ denotes the direction along a Ir-Ir given link: 

\begin{gather*}
\bm{D}_{ij} =  (0, 0, D)
      \qquad
      \bm{\mathcal{A}}_{ij} =  \begin{bmatrix}
   \mathcal{A}_{xx'} & 0 & 0 \\
   0 & \mathcal{A}_{yy'} & 0 \\
   0 & 0 & -\mathcal{A}_{xx'}-\mathcal{A}_{yy'} \\
      \end{bmatrix}
\label{matrices}
\end{gather*}

As shown in Ref.~\cite{Nguyen2021}, the latter term $\bm{\mathcal{A}}_{ij}$ is an order of magnitude smaller than $J$ and $\bm{D}_{ij}$ in \airo. Moreover it was shown to has little influence over $J$ and $D$ when included, i.e. that changes in these values are less than $10$~\%~\cite{Nguyen2021}. Therefore, to simplify our calculations, we restricted the Hamiltonian with the first two terms $J$ and $\bm{D}_{ij}$ only.

We thus calculated the magnon spectra of \tbiro\ with SpinW~\cite{Toth2015} by calculating the dynamical structure factor $S(\bm{Q}, \omega)$ = $S^{xx}(\bm{Q}, \omega) + S^{yy}(\bm{Q}, \omega)+S^{zz}(\bm{Q}, \omega)$ with the above Hamiltonian and starting from an all-in--all-out magnetic ground state. In order to fit our data, i.e. the experimental energy dispersion, and thus to extract $J$ and $D = |\bm{D}_{ij}|$, we used a heuristic method using the \textit{fitspec} function of SpinW~\cite{Toth2015}. 

\begin{figure}[!h]
\centering{\includegraphics[width=0.5\linewidth]{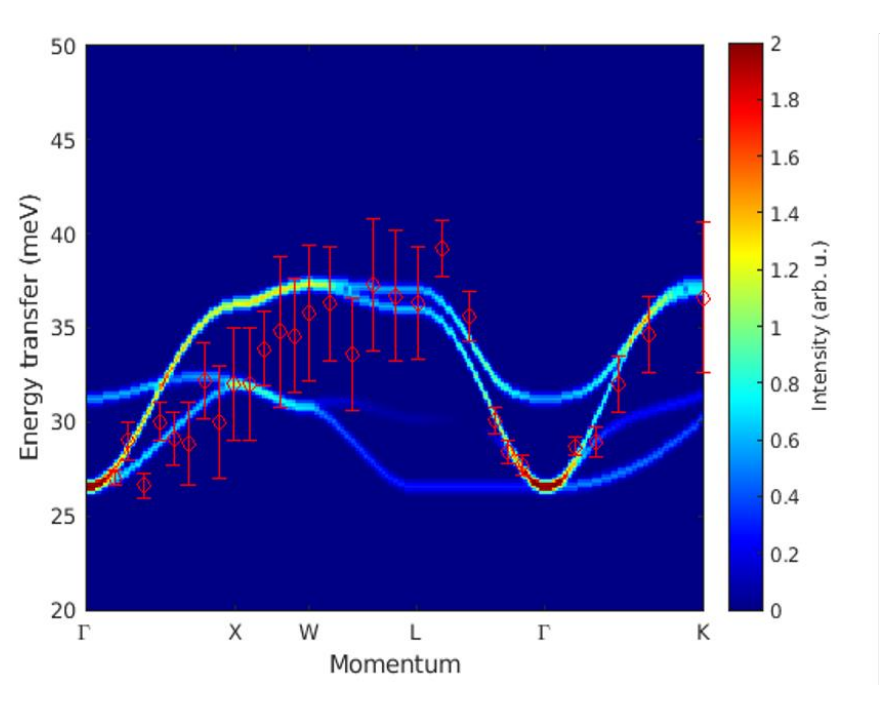}}
\caption{Calculated magnon spectra with $J = 16.2$~meV and $D = 5.2$~meV. Those values were obtained from fit of the experimental magnon dispersion (red open circles).}  
\label{Figure5}
\end{figure}

\begin{figure}[!h]
\centering{\includegraphics[width=0.80\linewidth]{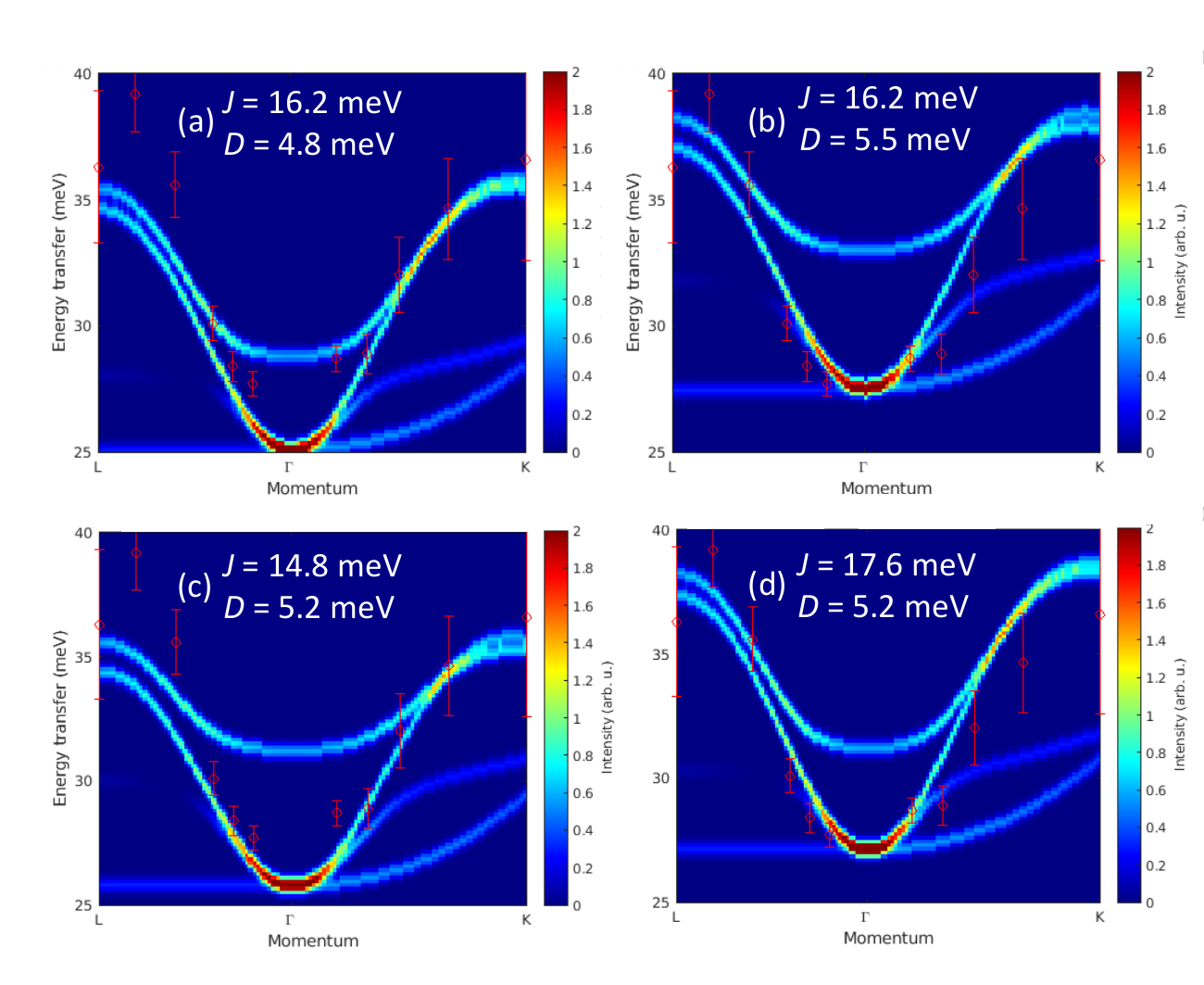}}
\caption{Calculated magnon spectra along $\rm{L} - \Gamma - \rm{K}$: (a-b) at fixed $J = 16.2$~meV and for $D = 4.8$~meV and $D = 5.5$~meV. (c-d) at fixed $D = 5.2$~meV for $J = 14.8$~meV and $J = 17.6$~meV. The red open circles denote the experimental magnon dispersion.}  
\label{Figure6}
\end{figure}

Fig.~\ref{Figure5} shows the calculated magnon spectra using the parameters $J = 16.2$~meV and $D = 5.2$~meV obtained from the best fit of the experimental magnon dispersion. In order to find the error bars of this set of parameters, we assumed that both parameters are independent and varied one parameter while keeping the other fixed until the calculated spectra does not match the experimental dispersion, in particular along $\Gamma-\rm{L}$ and $\Gamma-\rm{K}$ where experimental errors bars are minimal close to the $\Gamma$ point. Fig.~\ref{Figure6} shows such calculations. 

From that we estimate $J = 16.2(9)$~meV and $D = 5.2(3)$~meV leading to a ratio $D/J = 0.32(3)$.  

\FloatBarrier

\bibliography{supmatbib.bib}